\documentclass[cameraready]{Interspeech}

\usepackage{amsmath,amssymb}
\usepackage{graphicx}
\usepackage{booktabs}
\usepackage{multirow}
\usepackage{comment}
\usepackage{amsmath}
\usepackage{amssymb}
\usepackage[table]{xcolor}
\usepackage{placeins}
\usepackage{float}
\usepackage{pifont}
\usepackage{amsmath}
\usepackage{graphicx}
\title{Attention-Guided Reliability Scaling for Contrastive Decoding in Robust Audio-Visual Speech Recognition}

\author[affiliation={1}]{YoungChae}{Kim}
\author[affiliation={2}]{Da-Hee}{Yang}
\author[affiliation={1,2}, correspondingauthor]{Joon-Hyuk}{Chang}

\address{
    $^1$ Dept. Artificial Intelligence, Hanyang University, Seoul, Republic of Korea\\
    $^2$ Dept. Electronic and Electrical Engineering, Hanyang University, Seoul, Republic of Korea
}

\email{\{yc0604,douxi15,jchang\}@hanyang.ac.kr}


\begin{document}
\maketitle

\begin{abstract}

Large language model (LLM)-based audio-visual speech recognition (AVSR) systems are robust under noise. Contrastive decoding (CD), originally introduced to stabilize LLM generation by contrasting a weaker model against a stronger one at inference time, adjusts predictions without additional training. In this work, we apply CD to AVSR by contrasting audio-only conditioning with full audio-visual conditioning within the same underlying model. However, using a fixed contrastive strength introduces a trade-off across noise levels: stronger intervention helps under severe noise but may over-correct reliable predictions in clean conditions. We propose reliability-aware scaling of CD for AVSR. Instead of using a fixed strength, we adaptively modulate the contrastive influence at each token based on reliability signals derived from attention dynamics and inter-model predictive divergence. Experiments on LRS3 show consistent improvements across clean and low-SNR conditions.

\end{abstract}


\section{Introduction}

Existing approaches to improving noise robustness in Audio-Visual Speech Recognition (AVSR) primarily rely on structural modifications\cite{avhubert,autoavsr}, such as increasing the weight of visual features in the decoder or introducing dynamic gating modules at the decoding stage\cite{deepavsr,modalityattention}. While effective, these methods require additional fine-tuning\cite{whisperflamingo} and increase model complexity\cite{vcafe}, resulting in higher training costs and reduced deployment efficiency\cite{unifiedspeech,branchformeravsr}. Recent advances in integrating large language models (LLMs) into AVSR systems have demonstrated strong recognition performance\cite{llamavsr}. By combining the strong linguistic modeling capability of LLMs with audio-visual integration, these systems show improved robustness under noisy conditions\cite{matryoshkaavsr,mmsllama}. However, like prior AVSR systems, they may still be affected by corrupted acoustic inputs due to over-reliance on the audio modality under unreliable acoustic conditions\cite{modalityspecific,videotemporal}.





To mitigate this limitation without modifying model parameters, we introduce contrastive decoding (CD) into the AVSR setting as a training-free inference-time strategy. CD operates directly in log-probability space by contrasting predictions obtained under two conditioning settings\cite{contrastivedecoding,cdimprovesresoning,dola}. In our formulation, we use the same underlying LLM-based AVSR model under two conditioning settings: full audio-visual input for the Expert and audio-only input, obtained by omitting video embeddings, for the Amateur. This avoids architectural mismatch in contrastive comparison while enabling training-free decoding-time intervention. At each decoding step, we subtract the Amateur’s token-level log-probabilities from those of the Expert to attenuate acoustic bias. However, naively applying CD with a fixed contrastive weight across all decoding steps can introduce a critical limitation\cite{adaptivecd,adacad}. Acoustic reliability varies dynamically at the token level depending on signal-to-noise ratio (SNR) \cite{overconfidence,routergated}, yet static intervention treats all tokens uniformly. As a result, aggressive contrast may improve robustness in severe noise but unnecessarily distort reliable predictions in cleaner conditions, creating a fundamental trade-off between noise robustness and clean-speech preservation. Consequently, static CD fails to adapt to token-level reliability variations under real-world SNR fluctuations. 



To overcome this limitation, we propose attention-guided reliability scaling for contrastive decoding. Instead of applying a fixed contrastive weight, we dynamically modulate the intervention strength ($\lambda$) at each token using reliability signals derived from attention behavior and inter-model predictive divergence. By scaling contrastive strength according to acoustic reliability, our formulation stabilizes contrastive decoding under fluctuating noise conditions, yielding consistent performance improvements from clean to severely noisy environments—without requiring additional training or architectural modifications. The main contributions of this study are listed as follows:
\begin{itemize}
    \item We introduce \textbf{attention-guided reliability scaling} for contrastive decoding in LLM-based AVSR, providing a \textbf{training-free, inference-time} mechanism that adaptively modulates contrastive strength without architectural modification or additional fine-tuning.
    \item We analyze the limitations of fixed-weight contrastive decoding, showing that uniform intervention fails to account for token-level acoustic reliability and induces a robustness–clean-speech trade-off.
    \item We demonstrate through experiments that reliability-scaled contrastive decoding achieves consistent gains from clean to severely noisy conditions and generalizes across model scales and evaluation domains.
\end{itemize}

\begin{figure*}[t]
    \centering
    \includegraphics[width=0.95\textwidth]{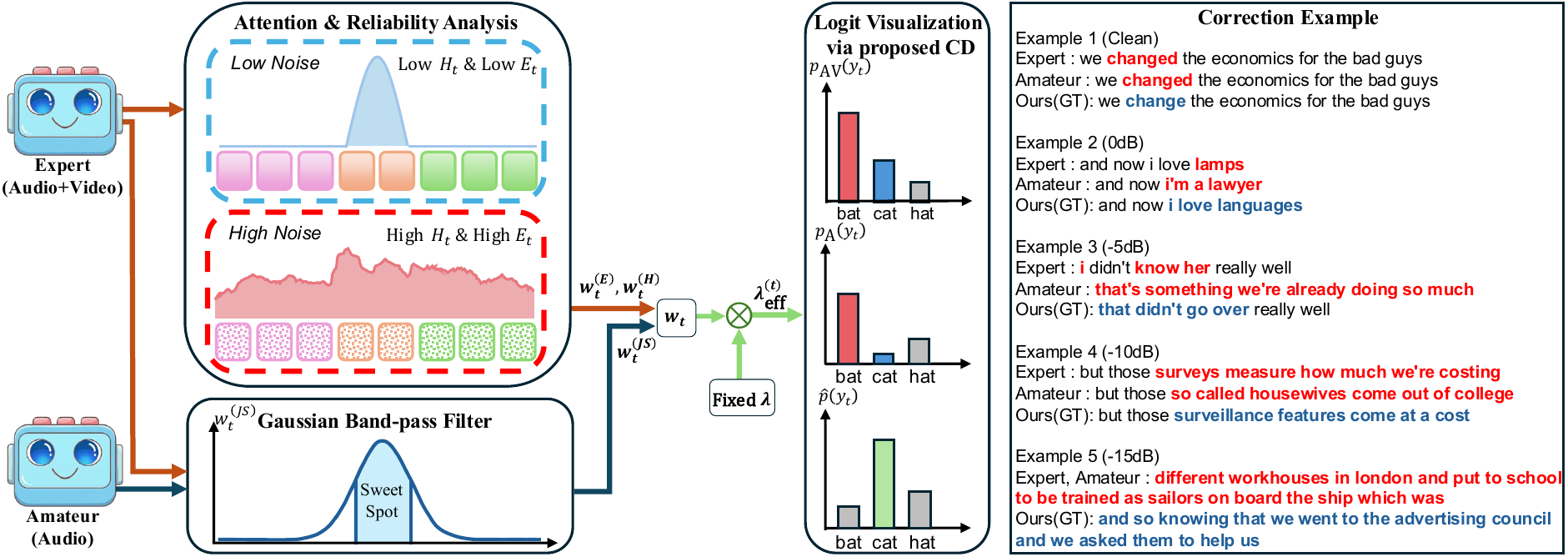} 
    \caption{Overall architecture of the proposed framework. Attention-based reliability signals and a JS divergence filter produce a token-level gate $w_t$, which rescales the contrastive weight $\lambda$ into the effective intervention strength $\lambda_t^{(\mathrm{eff})}$. The right panel shows correction examples under noisy conditions; red indicates errors and blue indicates correct predictions. The \textit{Ours} example is a sample whose prediction matches the ground truth (GT).}
    \label{fig:archi}
\end{figure*}

\section{Method}
\label{sec:method}


\subsection{Problem Definition and Motivation}

Contrastive decoding (CD) is a decoding-time strategy 
that contrasts token-level predictions under 
two different conditioning settings to suppress hallucinations \cite{contrastivedecoding,audioawaredecoding}. 
It adjusts the token log-probabilities with a contrastive weight $\lambda$ as:

\begin{equation}
\tilde{\ell}_t(y)
=
(1+\lambda)\log p_t^{\text{with}}(y)
-
\lambda \log p_t^{\text{without}}(y).
\end{equation}

However, applying a fixed contrastive strength $\lambda$ 
uniformly across all tokens is suboptimal in noisy AVSR settings. 
Audio attention, acoustic reliability, and inter-setting divergence 
fluctuate significantly at the token level. As a result, a static $\lambda$ cannot flexibly adapt to varying noise levels, leading to miscalibrated contrastive intervention across acoustic conditions.
To address this limitation, we introduce a token-level soft-gating 
mechanism that dynamically scales the contrastive strength.



In the AVSR decoding step $t$, we use the same underlying AVSR model under two different conditioning settings: 
the full audio-visual input as the Expert, and the audio-only input (with video embeddings omitted) as the Amateur. 
The effective contrastive strength is defined as
$\lambda_{\text{eff}}^{(t)} = w_t \cdot \lambda$, 
where $w_t \in [0,1]$ is a token-specific dynamic weight. 
This results in:

\begin{equation}
\log \hat{p}(y_t)
=
(1+\lambda_{\text{eff}}^{(t)})
\log p_{\text{AV}}(y_t)
-
\lambda_{\text{eff}}^{(t)}
\log p_{\text{A}}(y_t).
\end{equation}

Depending on the value of $w_t$, the mechanism gracefully interpolates between states: 
$w_t = 0$ yields pure AVSR decoding, 
$w_t = 1$ applies the standard full CD, 
and $0 < w_t < 1$ provides continuous, selective intervention.

\subsection{Attention Extraction and Reliability Metrics}
The dynamic weight $w_t$ is determined by the multiplicative fusion of three metrics representing different facets of model reliability: $w_t = w_t^{(E)} \cdot w_t^{(H)} \cdot w_t^{(JS)}$. We use multiplicative fusion to enforce conservative gating: CD activates only when all reliability signals are satisfied. The attention weights required for these metrics are extracted based on the structural properties of LLM-based AVSR model.

This architecture processes multiple modalities by concatenating them into a single embedding sequence. The input sequence layout follows the order: [BOS] [audio] $a_1 \dots a_{N_a}$ [/audio] [video] $v_1 \dots v_{N_v}$ [/video] [prompt] [text]. Because the audio feature tokens occupy a fixed index range from $s_a$ to $e_a$, i.e., $i \in [s_a, e_a)$, the number of audio tokens is defined as $N_a = e_a - s_a$. Based on this architecture, our attention analysis strictly adheres to three principles: we extract the weights assigned to the entire sequence by (1) the \textbf{last token ($T$)} currently being generated, at (2) the \textbf{last Transformer layer} of the model, and we (3) aggregate and average the values across \textbf{all attention heads ($\mathcal{H}$)} to prevent bias from any single head. We analyze the attention of the last decoding token at the final Transformer layer, as it directly contributes to the next-token probability distribution and therefore most faithfully reflects modality interaction during inference.


\subsubsection{Relative Audio Energy ($E_t$)}
This metric measures how much attention the current token allocates to the audio region. Let $\alpha_{T,i}^{(h)}$ be the attention weight assigned to position $i$ by head $h$. The energy is calculated as:

\begin{equation}
E_t = \frac{1}{|\mathcal{H}|} \sum_{h \in \mathcal{H}} \sum_{i=s_a}^{e_a-1} \alpha_{T,i}^{(h)}
\label{eq:energy}
\end{equation}

Since the absolute volume of audio attention fluctuates drastically depending on the overall SNR level, relying on absolute attention values fails to flexibly adapt to varying noise conditions. To overcome this, we construct an SNR-invariant, self-calibrating gate utilizing the running mean of the current utterance, $\bar{E}_t = \frac{1}{t}\sum_{s=1}^{t}E_s$:

\begin{equation}
w_t^{(E)} = \sigma \left( \beta_E \left( \frac{E_t}{\bar{E}_t} - 1 \right) \right)
\label{eq:energy_gate}
\end{equation}

At $t=1$, the energy gate yields a neutral weight ($w_1^{(E)}=0.5$), deferring early intervention decisions to complementary metrics.
While early tokens may show larger relative deviations from the running mean, the multiplicative formulation ensures that isolated energy variations do not disproportionately amplify the overall contrastive strength. 


This formulation addresses potential over-reliance on the audio modality under degraded conditions. When the acoustic signal is corrupted, the model may continue allocating disproportionate attention to the audio stream. 
By increasing the CD strength as the current audio attention $E_t$ rises above its running average $\overline{E}_t$, the mechanism reduces the tendency to over-rely on potentially corrupted acoustic cues and promotes greater use of complementary visual evidence.

The subtraction of 1 centers the ratio at zero, yielding a neutral gate value when $E_t = \bar{E}_t$, while $\beta_E$ controls the sensitivity of the sigmoid to deviations from this equilibrium. 

\subsubsection{Audio Entropy ($H_t$)}
This measures the uncertainty of the acoustic signal through the dispersion of attention within the audio region. Since multi-head attention is designed to capture diverse patterns across different heads, averaging the attention weights before calculating entropy would artificially inflate the uncertainty due to Jensen's inequality\cite{statproofbook}. Therefore, we calculate the entropy \textit{independently for each head} and then average the results. 
We first re-normalize the audio-specific attention for each head such that their sum equals 1: $\tilde{\alpha}_{T,i}^{(h)} = \alpha_{T,i}^{(h)} / \sum_{j=s_a}^{e_a-1}\alpha_{T,j}^{(h)}$. Then, to obtain a length-invariant metric strictly bounded within $[0, 1]$, we divide by the maximum possible entropy, $\log N_a$:
\begin{equation}
H_t = \frac{1}{|\mathcal{H}|} \sum_{h \in \mathcal{H}} \left( \frac{-\sum_{i=s_a}^{e_a-1} \tilde{\alpha}_{T,i}^{(h)}\log\tilde{\alpha}_{T,i}^{(h)}}{\log N_a} \right) 
\end{equation}
where $N_a = e_a - s_a$. A low entropy indicates concentrated attention on specific audio tokens (clear signal), whereas a high entropy implies dispersed attention (noisy or uncertain signal). The corresponding entropy gate is defined as:
\begin{equation}
w_t^{(H)} = \sigma\bigl(\beta_H(H_t - 0.5)\bigr) 
\end{equation}

Since $H_t$ is normalized to $[0,1]$, the midpoint $0.5$ serves as a neutral uncertainty pivot. The hyperparameter $\beta_H$ smoothly controls the intervention sensitivity around this boundary.

\subsubsection{Jensen–Shannon Divergence ($JS_t$)}
This metric quantifies the predictive disagreement between the Expert (audio-visual conditioning) and Amateur (audio-only conditioning) settings at decoding step $t$. Let $p_t^{\text{av}}$ and $p_t^{\text{a}}$ denote their predictive distributions, respectively. We compute the normalized Jensen–Shannon divergence as:
\begin{equation}
JS_t = JS(p_t^{\text{av}} \parallel p_t^{\text{a}}).
\end{equation}

We normalize the divergence by $\ln 2$ so that $JS_t \in [0,1]$. While contrastive intervention is generally beneficial under low-SNR conditions, blindly applying it during extreme inter-model disagreement can be unstable. As shown in Eq.~(2), when the Amateur distribution collapses—assigning near-zero probabilities to a large portion of the vocabulary—the term $\log p_{\text{A}}(y_t)$ becomes strongly negative for many tokens. Consequently, the subtraction term $-\lambda_{\text{eff}}^{(t)} \log p_{\text{A}}(y_t)$ injects large positive offsets across many tokens simultaneously. Importantly, this effect is token-dependent rather than a constant shift, as $\log p_{\text{A}}(y_t)$ varies across candidate tokens. Although the correct token may also receive amplification, the relative margins between plausible and implausible tokens can shrink unpredictably.


We refer to this phenomenon as \textbf{rank distortion}, where extreme disagreement may cause the contrastive term to introduce excessive token-wise offsets that destabilize the Expert's probability structure. To emphasize contrastive intervention under informative and reliable disagreement, we apply a Gaussian filter function:
\begin{equation}
w_t^{(JS)} 
=
\exp \left(
-\frac{(JS_t - \mu_{\text{sweet}})^2}{2\sigma_{JS}^2}
\right).
\end{equation}

This function emphasizes contrastive intervention when the disagreement is informative (near $\mu_{\text{sweet}}$), while suppressing it when the two distributions are nearly identical ($JS_t \approx 0$) or excessively divergent ($JS_t \to 1$).

\begin{table*}[t]
\centering
\small
\caption{ Performance comparison under different SNR conditions on LRS3 (in-domain) and LRS2 (OOD) test sets. Noise is injected using MUSAN at 0, -5, -10, and -15 dB. “A+V/A” denotes the WER obtained under full audio-visual conditioning (Expert) and audio-only conditioning (Amateur), respectively. Relative improvements are computed with respect to the AVSR baseline as $(\text{A+V} - \text{Ours}) / \text{A+V}$, and the reported average improvement is calculated across all SNR conditions.}
\resizebox{\textwidth}{!}{
\begin{tabular}{c c c c c c c c c}
\hline
Model(Size) & Test Domain & Method & Clean & 0 dB & -5 dB & -10 dB & -15 dB & Avg. Improvement \\
\hline
\multirow{4}{*}{Llama-AVSR (8B)}
& \multirow{2}{*}{LRS3}
& A+V/A & 0.0095/0.01 & 0.0367/0.0663 & 0.0945/0.2584 & 0.2358/0.7573 & 0.3723/1.0377 & - \\
& & Ours & \textbf{0.0082} (\textbf{+13.68\%}) & \textbf{0.033} (\textbf{+10.08\%}) & \textbf{0.0866} (\textbf{+8.36\%}) & \textbf{0.2167} (\textbf{+8.10\%}) & \textbf{0.3369} (\textbf{+9.51\%}) & \textbf{+9.95\%} \\
\cline{2-9}
& \multirow{2}{*}{LRS2}
& A+V/A & 0.0536/0.0643 & 0.0983/0.1527 & 0.1937/0.411 & 0.361/0.8575 & 0.4911/1.0739 & - \\
& & Ours & \textbf{0.0465} (\textbf{+13.25\%}) & \textbf{0.0878} (\textbf{+10.68\%}) & \textbf{0.1814} (\textbf{+6.35\%}) & \textbf{0.338} (\textbf{+6.37\%}) & \textbf{0.4518} (\textbf{+8.00\%}) & \textbf{+8.93\%} \\
\hline
\multirow{4}{*}{Omni-AVSR (1B)}
& \multirow{2}{*}{LRS3}
& A+V/A & 0.0142/0.0132 & 0.0495/0.0936 & 0.1175/0.3164 & 0.2509/0.803 & 0.3305/1.0441 & - \\
& & Ours & \textbf{0.0111} (\textbf{+21.83\%}) & \textbf{0.0468} (\textbf{+5.45\%}) & \textbf{0.1108} (\textbf{+5.70\%}) & \textbf{0.2306} (\textbf{+8.09\%}) & \textbf{0.3053} (\textbf{+7.62\%}) & \textbf{+9.74\%} \\
\cline{2-9}
& \multirow{2}{*}{LRS2}
& A+V/A & 0.0733/0.0832 & 0.1425/0.2111 & 0.2488/0.4967 & 0.3869/0.9287 & 0.4734/1.0952 & - \\
& & Ours & \textbf{0.0646} (\textbf{+11.87\%}) & \textbf{0.1344} (\textbf{+5.68\%}) & \textbf{0.2356} (\textbf{+5.31\%}) & \textbf{0.3607} (\textbf{+6.77\%}) & \textbf{0.456} (\textbf{+3.68\%}) & \textbf{+6.66\%} \\
\hline
\multirow{4}{*}{QWEN-AVSR (0.5B)}
& \multirow{2}{*}{LRS3}
& A+V/A & 0.0163/0.0185 & 0.0601/0.1033 & 0.1364/0.3298 & 0.2716/0.8048 & 0.3309/1.0476 & - \\
& & Ours 
& \textbf{0.0151} (\textbf{+7.36\%}) 
& \textbf{0.0572} (\textbf{+4.83\%}) 
& \textbf{0.1285} (\textbf{+5.79\%}) 
& \textbf{0.2559} (\textbf{+5.78\%}) 
& \textbf{0.3200} (\textbf{+3.29\%}) 
& \textbf{+5.41\%} \\
\cline{2-9}
& \multirow{2}{*}{LRS2}
& A+V/A & 0.0803/0.1164 & 0.1599/0.2386 & 0.2541/0.5083 & 0.3878/0.9162 & 0.4562/1.0802 & - \\
& & Ours 
& \textbf{0.0758} (\textbf{+5.60\%}) 
& \textbf{0.1524} (\textbf{+4.69\%}) 
& \textbf{0.2432} (\textbf{+4.29\%}) 
& \textbf{0.3675} (\textbf{+5.23\%}) 
& \textbf{0.4431} (\textbf{+2.87\%}) 
& \textbf{+4.54\%} \\
\hline
\end{tabular}%
}
\label{tab:main_results}
\end{table*}

\section{Experiments and Results}
\label{sec:experimental_setup}

\subsection{Datasets and Metrics}
We evaluate our method on the LRS3 dataset, which comprises 433 hours of training data \cite{lrs3}. To assess robustness under acoustic degradation, we artificially corrupt the standard test set (1,327 utterances) using the MUSAN dataset\cite{musan}. Specifically, we randomly mix noise, speech, and music in equal proportions and inject them into the original audio signals to achieve four SNR levels: 0 dB, -5 dB, -10 dB, -15 dB. Transcription quality is evaluated using the Word Error Rate (WER). In addition to in-domain evaluation on LRS3, we evaluate out-of-distribution(OOD) generalization on LRS2\cite{lrs2}, which differs in speaker distribution and recording conditions, applying the same MUSAN-based corruption protocol to maintain consistent SNR settings.

\begin{figure}[H] 
    \centering
    \includegraphics[width=\columnwidth]{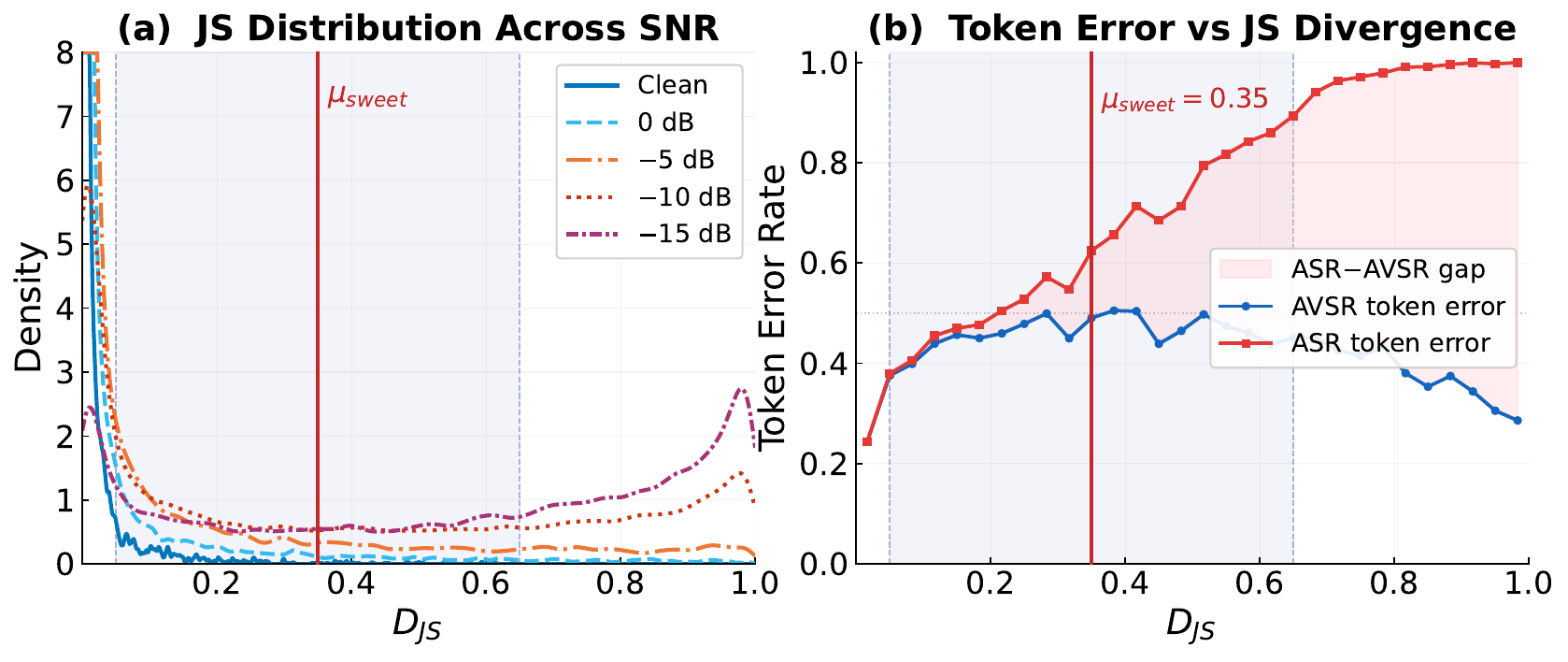}
    \caption{Token-level JS divergence analysis across SNR levels. The Gaussian filter center ($\mu_{\text{sweet}}$) is selected based on validation-set analysis to capture the onset of ASR unreliability. Here, AVSR denotes audio–visual conditioning, whereas ASR denotes audio-only conditioning.}
    \label{fig:js_analysis}
\end{figure}

Figure~\ref{fig:js_analysis}(a) shows that as SNR decreases, the JS distribution progressively shifts rightward, and at $-15$~dB it exhibits a bimodal structure with peaks near $JS \approx 0$ and $JS \approx 1$. 
This indicates the coexistence of tokens where the two models easily agree and tokens whose predictions are dominated by collapse under extreme noise. Figure~\ref{fig:js_analysis}(b) further shows that ASR token error increases sharply with JS, whereas AVSR remains relatively stable over a range of disagreement levels. In this regime, the AVSR predictions are not yet fully collapsed into random noise, but rather reflect a transitional phase where ASR reliability begins to degrade.
These observations motivate centering the Gaussian filter around the onset of ASR unreliability ($\mu_{\text{sweet}} = 0.35$), enabling selective intervention in informative disagreement regions while suppressing collapse-dominated extremes. 
Moreover, because the final weight is multiplicatively combined as $w_t = w_t^{(E)} \cdot w_t^{(H)} \cdot w_t^{(JS)}$, CD is activated only when energy, entropy, and JS signals jointly indicate unreliability, thereby preventing excessive intervention. 

\subsection{Model Architectures and Training Setup}
We evaluate our framework on three LLM-based AVSR systems with different scales: \textbf{Llama-AVSR} (Llama-3.1-8B), \textbf{Omni-AVSR} (Llama-3.2-1B), and \textbf{Qwen-AVSR} (Qwen2.5-0.5B).

\textbf{Architecture Details.}
Across all settings, visual features are extracted using a frozen AV-HuBERT-Large encoder\cite{avhubert}, and audio features are extracted using a frozen Whisper\cite{whisper} encoder: \textit{Whisper-Small} for Omni-AVSR\cite{omniavsr} and Qwen-based AVSR variant (Qwen-AVSR)\cite{qwen25}, and \textit{Whisper-Medium} for Llama-AVSR\cite{llamavsr}. The base LLM and encoders remain frozen, with LoRA\cite{lora} applied to projection layers. The architecture also supports audio-only conditioning, which is used to compute Jensen-Shannon divergence.

\textbf{Training Data.}
Llama-AVSR and Omni-AVSR use checkpoints trained on LRS3+VoxCeleb2\cite{voxceleb2} (1,756 h), while Qwen-AVSR uses a checkpoint trained only on LRS3 (433 h) to evaluate performance in a lower-resource setting.


\subsection{Implementation Details}
All experiments were conducted on NVIDIA A100 GPUs. Under batch size 1 with greedy decoding, the proposed method increases per-utterance decoding latency from 1577.9 ms for standard AVSR decoding (without CD) to 1714.3 ms (+136.4 ms, +8.6\% overhead), indicating a modest computational cost.
The base contrastive weight $\lambda$ is fixed at $0.3$. 
At each decoding step $t$, the intervention weight $w_t$ is computed based on the reliability signals defined in Section~\ref{sec:method}. 
The gating functions are parameterized by sigmoid sharpness ($\beta_E=10.0, \beta_H=10.0$) and Gaussian filter parameters ($\mu_{\text{sweet}}=0.35, \sigma_{JS}=0.15$). 
These hyperparameters are tuned on a held-out validation subset and kept fixed during test-time evaluation.
We observed stable performance within $\beta \in [5,15]$ and $\mu \in [0.3,0.4]$, indicating that the proposed method is not highly sensitive to these settings.

\subsection{Main Results}
\label{sec:results}
Table~\ref{tab:main_results} demonstrates that the proposed method generalizes consistently across model scales and evaluation domains. 
The same decoding rule transfers reliably across models regardless of parameter size, maintaining consistent gains on LRS3 and LRS2. Notably, our method improves clean-condition WER across all model sizes and maintains gains under noisy conditions, indicating improved robustness while also enhancing clean-speech accuracy.

\subsection{Analysis and Ablation}
\label{sec:adaptability}
\begin{table}[!ht]
\centering
\caption{Performance of CD under different fixed $\lambda$ values without reliability gating. Gray shading marks the best result among static fixed-$\lambda$ CD settings for each SNR condition.}
\resizebox{\columnwidth}{!}{
\begin{tabular}{l c c c c c}
\hline
contrastive weight $\lambda$ & Clean & 0\,dB & -5\,dB & -10\,dB & -15\,dB \\
\hline
Baseline (No CD) & 0.0095 & 0.0367 & 0.0945 & 0.2358 & 0.3723 \\
\hline
$\lambda=0.1$ & \cellcolor{gray!20}\textbf{0.0081} & \cellcolor{gray!20}0.0332 & \cellcolor{gray!20}0.0889 & 0.2187 & 0.3385 \\
$\lambda=0.2$ & 0.0090 & 0.0353 & 0.0896 & 0.2160 & 0.3239 \\
$\lambda=0.3$ & 0.0101 & 0.0372 & 0.0968 & 0.2143 & 0.3180 \\
$\lambda=0.4$ & 0.0114 & 0.0410 & 0.1011 & \cellcolor{gray!20}\textbf{0.2030} & \cellcolor{gray!20}\textbf{0.3150} \\
$\lambda=0.5$ & 0.0130 & 0.0463 & 0.1091 & 0.2237 & 0.3196 \\
\hline
\textbf{Ours (Adaptive)} & 0.0082 & \textbf{0.0330} & \textbf{0.0866} & 0.2167 & 0.3369 \\
\hline
\end{tabular}
}
\label{tab:lambda_sweep}
\end{table}

Table~\ref{tab:lambda_sweep} highlights a fundamental limitation of fixed-weight contrastive decoding. The optimal contrastive weight is highly condition-dependent, as stronger contrast improves robustness under severe noise but may penalize reliable predictions in cleaner conditions. 
In practical deployment, however, the SNR of incoming speech is unknown and varies dynamically across utterances and tokens, making manual tuning of $\lambda$ for specific noise conditions infeasible and limiting the practicality of fixed-weight CD. We therefore set the base contrastive weight to $\lambda = 0.3$, which achieved the best overall performance across the explored range.

\begin{table}[!ht]
\centering
\caption{Ablation study on Llama-AVSR (8B) under different SNR conditions. For all experiments involving CD, the parameter $\lambda$ is fixed at 0.3.}
\resizebox{\columnwidth}{!}{
\begin{tabular}{l c c c c c c c c}
\hline
Method & $E_t$ & $H_t$ & $JS_t$ & Clean & 0 dB & -5 dB & -10 dB & -15 dB \\
\hline
Baseline & - & - & - & 0.0095 & 0.0367 & 0.0945 & 0.2358 & 0.3723 \\
w/ CD + $E_t$ & \ding{51} &  &  & 0.0086 & 0.0339 & 0.0883 & 0.2189 & 0.3332 \\
w/ CD + $H_t$ &  & \ding{51} &  & 0.0095 & 0.0364 & 0.0952 & \textbf{0.2147} & \textbf{0.3202} \\
w/ CD + $JS_t$ &  &  & \ding{51} & \textbf{0.0082} & 0.0346 & 0.0883 & 0.218 & 0.322 \\
Ours & \ding{51} & \ding{51} & \ding{51} & \textbf{0.0082} & \textbf{0.033} & \textbf{0.0866} & 0.2167 & 0.3369 \\
\hline
\end{tabular}
}
\label{tab:ablation}
\end{table}

Table~\ref{tab:ablation} illustrates the complementary behaviors of the proposed reliability cues. The JS divergence primarily stabilizes decoding in clean and mildly noisy conditions, such as 0 and -5~dB, by mitigating excessive contrastive influence from unreliable audio-only predictions. In contrast, $E_t$ and $H_t$ contribute more noticeably under moderate noise levels, such as -10 and -15~dB, where acoustic reliability gradually degrades. When combined multiplicatively, these signals yield balanced performance across SNR conditions, indicating that conservative activation effectively stabilizes inference without aggressive contrastive intervention.

\section{Conclusion}
In this work, we proposed a reliability-aware contrastive decoding strategy that adaptively scales intervention strength using attention and divergence cues. Unlike static approaches that force a compromise between denoising capability and clean accuracy, our method simultaneously improves robustness under severe noise while preventing degradation in clean settings. As a plug-and-play, training-free solution, it offers a practical path to robust AVSR without typical performance trade-offs.

\section{Acknowledgement}
This work was supported by Institute of Information \& communications Technology Planning \& Evaluation (IITP)grant funded by the Korea government(MSIT) (No.RS-2020-II201373, Artificial Intelligence Graduate School Program(Hanyang University)).


\section{Generative AI Use Disclosure}
Generative AI tools were used for language polishing and proofreading of the manuscript.



\bibliographystyle{IEEEtran}
\bibliography{mybib}

\end{document}